\documentclass[useAMS,usenatbib]{mn2e}

\title[Li abundance in lower RGB stars]
  {{\sl Giants reveal what dwarfs conceal:}
  Li abundance in lower RGB stars as diagnostic of the 
  primordial Li
  \thanks{Based on observations collected at the ESO-VLT under programs 068.D-0546, 
  072.B-0585, 266.D-5655, 077.D-0018, 65.L-0165 and on data available in the 
  ELODIE archive. 
 These research has also made use of the SIMBAD database, operated at CDS, Strasbourg, France.
  }}

\author[Mucciarelli et al.]
  {A. Mucciarelli,$^1$ M. Salaris,$^2$  and P. Bonifacio$^3$
  \\
  $^1$Dipartimento di Astronomia, Universit\`a 
  degli Studi di Bologna, Via Ranzani, 1 - 40127, 
  Bologna, Italy
  \\
  $^2$Astrophysics Research Institute, Liverpool John Moores University, 
    12 Quays House, Birkenhead, CH41 1LD, United Kingdom  
   \\
   $^3$GEPI, Observatoire de Paris, CNRS, Univ. Paris Diderot, 92125, Meudon Cedex, France
}
\usepackage{graphicx}

\usepackage{color}
\pagerange{\pageref{firstpage}--\pageref{lastpage}} \pubyear{2002}

\def\LaTeX{L\kern-.36em\raise.3ex\hbox{a}\kern-.15em
    T\kern-.1667em\lower.7ex\hbox{E}\kern-.125emX}

\begin{document}

\label{firstpage}

\maketitle

\begin{abstract}
The discrepancy between cosmological Li abundance inferred from Population~II dwarf stars 
and that derived from Big Bang nucleosynthesis calculations is still far from being satisfactorily solved.
We investigated, as an alternative route, the use of Li abundances in Population~II lower red giant branch stars 
as empirical diagnostic of the cosmological Li. Both theory and observations 
suggest that the surface Li abundance in metal poor red giants after the completion of the 
first dredge-up and before the red giant branch bump, are significantly less sensitive 
to the efficiency of atomic diffusion, compared with dwarf stars.
The surface Li abundances in these objects --  after the dilution 
caused by the first dredge-up -- are predicted to be sensitive to the total Li content left in the star, i.e. they 
are affected only 
by the total amount of Li eventually burned during the previous main sequence phase.
Standard stellar models computed under different physical assumptions show that the inclusion of the 
atomic diffusion has an impact of about 0.07~dex in the determination of the primordial 
Li abundance --  much smaller than the case of metal poor main sequence-turn off
stars --  and it is basically unaffected by reasonable variations other parameters (overshooting, age, 
initial He abundance, mixing length).
We have determined from spectroscopy the surface Li content of 17 Halo lower red giant branch stars, 
in the metallicity range 
between [Fe/H]$\sim -$3.4 and $\sim -$1.4 dex, evolving before the extra-mixing episode that sets in at the 
red giant branch bump.
The initial Li (customarily taken as estimate of the cosmological Li abundance ${\rm A(Li)_0}$) has then 
been inferred by 
accounting for the difference between initial and post-dredge up Li abundances in the appropriate stellar models. 
It depends mainly on the $T_{eff}$ scale adopted in the spectroscopic
analysis, and is only weakly sensitive to the efficiency of atomic diffusion in the models, 
so long as one neglects Li destruction caused by the process competing with atomic diffusion. 
Our final ${\rm A(Li)_0}$ estimate spans a relatively narrow range, 
between 2.28 and 2.46~dex, and is $\sim$0.3--0.4~dex lower 
than predictions from Big Bang nucleosynthesis calculations.
These values of ${\rm A(Li)_0}$ are corroborated by the analysis of samples of red giants 
in the Galactic globular clusters NGC~6397, NGC~6752 and M4. 
Our result provides an independent quantitative estimate of the difference with the 
Big Bang value, and sets a very robust constraint for the physical processes invoked to resolve this discrepancy.

\end{abstract}


\section{Introduction}

The discovery by \citet{spite82,spite82b} of a uniform Li abundance in the atmospheres of 
main sequence (MS) Halo field stars with $T_{eff}$ above $\sim$5500 -- 5900~K and 
[Fe/H] below about $-$1.5~dex -- the so-called {\sl Spite Plateau} --  has been widely interpreted 
as a signature of the cosmological abundance of lithium, produced during the Big Bang nucleosynthesis (BBN). 
Adopting the standard notation A(Li)=log(n(Li)/n(H))+12, 
the {\sl Spite Plateau} value is between 2.1 and 2.4, 
depending mostly on the adopted $T_{eff}$ scale 
\citep[see, e.g.,][and references therein]{boni97, charb05, asplund06, boni07, aoki09,sbordone10}. 

Estimates of the cosmological baryon density obtained from 
the power spectrum of the cosmic microwave background fluctuations \citep{spergel07} 
combined with standard BBN calculations, 
predict however a cosmological ${\rm A(Li)_0}$=2.72$\pm$0.06 \citep{cyburt08}.
Taken at face value, the BBN A(Li) is higher 
-- by at least a factor of 2 -- than A(Li) measured in {\sl Spite Plateau} stars.

The interpretation of {\sl Spite Plateau} abundances in terms of the 
initial chemical composition of Halo stars needs however to take into account the possible 
effect of stellar evolution on the star surface chemical composition. 
Stellar models that include only convection as element transport mechanism predict for {\sl Spite Plateau} stars  
a negligible pre-MS Li depletion, and no changes of the  
surface abundances during the whole MS phase. With these assumptions, the observed A(Li)   
is essentially equal to the primordial value.

On the other hand, detailed stellar evolution calculations need to include additional transport mechanisms that 
follow from first principles. In low mass stars, atomic diffusion  -- i.e. the slow transport of chemicals 
due to temperature, pressure and abundance gradients -- 
is the main process that needs to be included, for taking diffusion into account, the agreement  
between solar models and constraints from helioseismology is greatly improved 
\citep[see, i.e.,][]{bahc97}. 
A general prediction of models including atomic diffusion is that Population II  MS stars in 
the {\sl Spite Plateau} region, 
with their shallow convective envelopes, are efficiently depleted of Li and metals over the MS lifetime 
\citep[see, e.g., the seminal paper by][]{del90}. Moreover, 
the surface Li depletion is a function of the star initial metallicity (and $T_{eff}$), because of the 
associated variation of the depth of the convective envelopes.  
These predictions are at odds with the observed constant 
(both as a function of [Fe/H] and $T_{eff}$)  
A(Li) along the plateau 
\citep[but see also][for the evidences of a meltdown of the Spite plateau in the very metal-poor regime]{sbordone10}. 
Inclusion of radiative levitation \citep{richard05} leaves the overall picture unchanged. 
For [Fe/H] below $\sim -$2.0~dex radiative levitation moderates the surface depletion of Li due 
to atomic diffusion but the general picture is not qualitatively modified, 
whereas at higher metallicities the effect on the surface Li is negligible
\footnote{One has to notice that in \citet{richard05} the term atomic diffusion includes 
also the process of radiative levitation}. Overall, the depletion of Li in {\sl Spite Plateau} stars 
due to atomic diffusion (plus radiative levitation) can reach several tenths of a dex at the lowest 
metallicity and/or higher $T_{eff}$.

For the models with diffusion (and radiative levitation) to display uniform A(Li) abundances 
along the {\sl Spite Plateau}, one needs to include some 
{\sl ad-hoc} turbulent mixing in the radiative zone below the convective boundary, 
that limits the settling of Li. \citet{richard05} include a specific  
parametrization of turbulent mixing, treated as a diffusive process, with a free parameter that fixes 
the values of the turbulent diffusion coefficient. A suitable  
calibration of this  parameter -- that enables also the transport of additional 
lithium down to burning temperatures -- 
produces a flat Li abundance along the plateau, with A(Li) depleted 
by about 0.4~dex with respect to the initial value \citep[see Fig.~ 8 in][]{richard05}.
This depletion, when applied to the range of A(Li) measured by different authors, can potentially 
bring into agreement the {\sl Spite Plateau} with BBN predictions.

\citet{piau} includes instead the effect of a rotationally induced tachocline mixing \citep{sz92} as 
employed in solar models \citep{brun} to improve the agreement between theoretical 
and observed sound speed. Also in this case mixing is treated as a diffusive process, and the author 
discusses the calibration of the associated free parameters and the dependence on the assumed 
rotation history. As a result, tachocline mixing -- that in plateau stars 
moderates the efficiency of atomic diffusion -- appears to be able to produce a uniform 
A(Li) along the {\sl Spite Plateau}, with a depletion of 0.2~dex compared to the initial value. 
Given the range of measured A(Li), the agreement with the BBN lithium abundance is still marginal at best. 

Similar estimates of primordial Li based on spectroscopy of MS stars in globular clusters are complicated 
by the fact that in many cases the observed A(Li) display spreads related to the presence of second 
generation stars \citep[see, e.g.,][for the case of 47~Tuc]{boni07,dorazi10}. 
The metal poor cluster NGC6397 is one object with apparently an essentially homogeneous initial Li abundance and  
\citet{korn06} spectroscopic measurements plus the models by \citet{richard05} with a suitably calibrated 
efficiency of their turbulent mixing parametrization -- that turns out to be different from the 
case of field halo dwarfs -- are able to 
reproduce the observed A(Li) (and [Fe/H]) measurements at the MS turn off and the base of the red giant branch (RGB), 
for an initial A(Li)=2.54. The models fail however to match the observed post dredge-up A(Li) on the lower RGB, 
before the RGB bump. A recent reanalysis by \citet{lind} have revised down this value, to A(Li)=2.46. 
A detailed comparison with the observed post dredge-up Li abundance is not presented. 
Moreover, \citet{lind} conclude: 
``We find that some turbulence, in a very limited efficiency-range, is indeed required to explain observations. 
These models'' \citep[including atomic diffusion plus turbulent mixing parametrized as in][]{richard05} 
``fail to reproduce the behaviour of Li abundance with effective temperature along 
the plateau, suggesting that a detailed understanding of the physics responsible for depletion is still lacking.''
Note also that the assumption of a 
different temperature scale leads to different results, in terms of match between observations and 
models, as pointed out by \citet{jonay}.

Another cluster that does not show signatures of a spread in the initial Li abundance is M4. 
\citet{muc} shows that models including diffusion plus turbulence with 
the same efficiency invoked for NGC6397, fail in case of M4. It appears that, in order to reproduce 
the Li abundances from the turn off to the lower RGB and satisfy at the same time the BBN constraint, 
for this cluster the turbulence has to reach deeper regions compared to NGC6397, 
region hot enough for some additional Li burning to occur. 
The trend of lithium abundance with effective temperature
along the sub-giant branch, found in M4  by \citet{muc} is similar
to that found by \citet{jonay} in NGC6397.

This brief summary of previous investigations shows how problematic, from a theoretical point of view,  
it is to determine the primordial Li abundance of {\sl Spite Plateau} stars. 
In this paper we investigate a complementary avenue to estimate the Halo primordial 
A(Li). We employ spectroscopy of Halo stars evolving along the lower RGB, defined 
-- following \citet{gratton00} -- as the portion of the RGB 
brighter than the luminosity corresponding to the completion of the first dredge-up 
and fainter than the RGB-bump. We will show that the effect of atomic diffusion on the 
surface Li abundances of these objects is much smaller than for {\sl Spite Plateau} stars. As a 
consequence, the observed A(Li) can be employed to 
set independent, very strong constraints on any additional physical process  
(i.e., turbulent mixing, preprocessing of Halo material, modifications to the BBN) 
eventually needed to reconcile these values with BBN predictions.

The paper is structured as follows. Section~2 analyzes the theoretical advantages of employing  
A(Li) measured in the atmosphere of lower RGB Halo stars to estimate their primordial Li abundance,  
while Sect.~3 discusses our selected sample of lower RGB halo objects and the derivation of 
A(Li) and [Fe/H]. Section~4 presents theoretical predictions for the surface Li abundance in 
Population II lower RGB stars --  from models including 
atomic diffusion and convection as element transport mechanisms -- 
and the derivation of the initial Li for our selected star sample, that 
provides an estimate of the Halo primordial Li abundance independent of the {\sl Spite Plateau}. 
Section 5 discusses the Li abundance in the lower RGB of the Galactic globulars NGC~6397, NGC~6752 and M4.
A summary and conclusions close the paper.

\section{Why lower RGB stars ?}

The surface Li abundance after the first dredge-up is essentially a consequence of 
the dilution due to the increased size of the convective envelope after the MS turn off 
(plus a minor contribution from Li burning in the deep layers of the fully mixed envelope in the 
most metal rich Halo stars). At the end of the MS phase, when the deepening convective boundary reaches 
layers where the Li-burning ($T_{burn}\sim$2.5 $10^6$~K) was efficient during the MS, 
the surface A(Li) begins to decrease. This depletion essentially ends when the convective envelope attains its 
maximum depth and the first dredge-up is complete. 
An important point to notice is that atomic diffusion during the MS produces a local maximum of the Li abundance  
in the radiative layers right below the convective envelope \citep[see, e.g.,][]{richard05}, for  
only a relatively small fraction of the envelope Li is transported deep enough to reach layers where it is 
eventually burned. 
As a consequence, models without and with diffusion -- fully efficient or even moderated by levitation or 
some turbulence that just mixes back material diffused from the envelope -- dilute similar 
amount of Li within the deepening convective region. In addition, the maximum size of the convective envelope is also 
weakly affected by diffusion -- this can be also inferred by the fact that the 
predicted RGB bump luminosities with and without diffusion are very similar \citep{cas97, mic10} -- and the resulting 
A(Li) abundances on the lower RGB are only slightly changed. 
This is in contrast with 
models for upper MS stars, where the effect of diffusion 
on the surface abundances can reach several tenths of dex \citep[see for example Fig.~3 in][]{muc}. 
Theoretical models also predict that along the RGB, after the completion of the dredge-up, atomic diffusion and levitation 
are not able to modify appreciably the surface abundance of Li and all other elements \citep{mic07, mic10}. 

The chemical abundance measurements in giant stars by \citet{gratton00}, \citet{spite05} and  \citet{lind}, suggest 
that any additional element transport along the RGB is very likely inefficient in this phase. On the other 
hand, stars evolved beyond the RGB bump display the effect of an additional mixing event, for which the so-called 
'thermohaline mixing' is nowadays the most popular candidate \citep[see, e.g.][and references therein]{charb}.

These considerations lead to the conclusion that in general there is a simpler 
relationship between initial and current 
surface Li abundances in lower RGB stars compared to {\sl Spite Plateau} objects. The basic reason is that, 
for a fixed initial Li abundance, lower RGB surface abundances are sensitive to the total amount of 
Li left in the star. In fact, after 
the effect of dilution is accounted for, they 
are affected only by the total amount of Li burned during the MS phase due to atomic diffusion   
plus possible additional element transport mechanisms -- if they are invoked to solve the discrepancy with BBN results.
On the other hand, the observed abundances along the {\sl Spite Plateau} are determined by 
the evolution of the rate of Li depletion from the convective envelope due just to diffusion 
or moderated/enhanced by additional turbulent processes. 
Combining the information obtained from Li in lower RGB stars with {\sl Spite Plateau} data, sets much stronger constraints 
on how these proposed 
additional mixing processes act between the base of the convective envelope and the deeper Li burning regions. 

In this study, we address the issue of the Halo primordial Li abundance by coupling predictions from 
standard RGB stellar models to the Li abundances measured on a sample of lower RGB Population II field stars. 
As discussed before, this will provide an independent quantitative estimate of the difference (if any) with the 
Big Bang value, and set a very robust constraint on the efficiency of additional physical processes invoked to resolve 
this discrepancy, complementary to the interpretation of the abundances along the {\sl Spite Plateau}.
If the solution of the discrepancy involves pre-processing during the Galaxy formation or modifications to the BBN 
\citep[see, e.g.,][]{piau06, cyburt10}, our analysis will provide a solid quantitative estimate of, respectively, 
the net amount of Li burned during the early Galaxy evolution, or exactly how much Li must be synthesized in the 
'revised' BBN.

To the best of our knowledge, analyses of the Li abundances measured along the RGB have been so far mainly aimed at 
testing the agreement between observed and predicted Li depletion after the first dredge-up, assuming a value 
for the initial A(Li), often equal to what measured along the plateau \citep{gp06}.

\section{Li abundances in lower RGB field Halo stars}

This section presents the observational data employed in our study, 
and the abundance analysis. Particular care is paid to the issue of the effective temperature scale 
and its influence on the derived values of A(Li).

\subsection{Dataset}
We have selected a sample of 17 metal-poor stars (with [Fe/H]$<$--1 dex), located 
in the part of the RGB fainter than the RGB bump magnitude level. The selection 
of this sample was done by cross checking data by \citet{cayrel01}, the SIMBAD 
website \footnote{http://simbad.u-strasbg.fr/simbad/}, 
previous works on chemical abundances in metal-poor giant stars \citep{burris, gratton00, johnson02} 
and the ESO and ELODIE archives, in order to select suitable 
high-resolution spectra, covering the Li line at 6708 $\mathring{A}$.

Spectra of 11 stars were retrieved from the ESO archive and observed with UVES \citep{dekker}, 
employing simultaneously different gratings. We used spectra obtained with the 
CD\#3 cross-disperser with a nominal spectral resolution of $\sim$54000 and a spectral coverage 
between $\sim$4760 and 6820 $\mathring{A}$, to measure Li and Fe abundances, and spectra obtained 
with the grating CD\#1 ($\sim$3050--3870 $\mathring{A}$) or CD\#2 ($\sim$3760--4980 $\mathring{A}$)
to measure C abundances and the $^{12}C/^{13}C$ isotopic ratios. 
For 6 stars we used spectra from the ELODIE archive \citep{moult04}, with a spectral resolution of 
$\sim$42000 and a coverage between $\sim$4000 and 6800 $\mathring{A}$. 

Table~\ref{dat} summarizes the main information for these stars (identification number of the 
corresponding catalog, adopted instrument, atmospherical parameters, abundances and 
SNR per pixel around the Li doublet).

\subsection{Abundance analysis}

The abundance analysis was performed by means of the codes 
SYNTHE \citep[to compute synthetic spectra,][]{K93_18,K05} 
and WIDTH \citep[that predicts the abundance of a given species by matching the observed and theoretical equivalent widths,][]{K93_13,castelli05}, 
coupled to the ATLAS9 model atmospheres \citep{K93_13,K05}. We used the Linux version
of all the codes \citep{sbordone04,sbordone05}. 
The ATLAS9 models employed in this work were computed with the new 
set of Opacity Distribution Function \citep{castelli04} and without the inclusion of the approximate 
overshooting in the calculation of the convective flux.

Iron abundances were obtained from line equivalent widths (EWs) using WIDTH; 
EWs were measured using the DAOSPEC code \citep{stetson}, that performs an automatic 
measurement of the EWs under the gaussian profile approximation.
The adopted linelist includes a hundredth of neutral iron lines whose oscillator 
strengths are from the recent compilation of accurate laboratory log~{\sl gf} 
by \citet{fuhr} and $\sim$15-20 single ionized iron lines by \citet{raa}. 
For each star we refine iteratively the linelist in order to include only lines 
predicted to be unblended from the inspection of synthetic spectra computed with the 
atmospherical parameters and metallicity of each target.
The resulted iron abundances were scaled to the solar iron abundance of 
7.52 \citep{caffau10}. Total uncertainties in the Fe abundances were computed by taking into 
account the internal error (obtained as the dispersion of the mean normalized to the 
root mean square of the number of used lines) and the errors related to the 
atmospherical parameters.

Li abundances were derived with the help of spectral synthesis, in order to 
take into account the hyperfine structure and the isotopic splitting of 
the Li resonance doublet at 6707.8 $\mathring{A}$. The linelist for the Li 
transitions is from \citet{yan95}. Corrections for departures from LTE are from 
\citet{carlsson94}. The total error in the A(Li) abundance is computed by taking into 
account the two main source of uncertainty: (i)~the error in the adopted $T_{eff}$, 
typically $\delta A(Li)/\delta T_{eff}\sim$0.09-0.10 dex per 100 K for a giant star 
(giant stars are slightly more sensitive to $T_{eff}$ with respect to the dwarf stars, 
that are sensitive at a level of $\delta A(Li)/\delta T_{eff}\sim$0.06-0.07 dex per 100 K); 
(ii)~the error in the fitting procedure, obtained by employing MonteCarlo simulations: 
for each star, we injected Poisson noise into the best-fit synthetic spectrum 
(in order to reproduce the observed SNR around the Li line)
and we repeated the analysis for 1000 MonteCarlo events. We estimated a 1$\sigma$ 
level from the resulting abundance distributions; due to the high 
SNR of the analyzed spectra ($>$100), the typical errors are lower than 0.03 dex.
The errors due to uncertainties on gravity and microturbulent velocity are negligible 
(of the order of 0.01~dex or less).

We measured also the isotopic ratio $^{12}C$/$^{13}C$ in order to robustly  assess 
whether the stars have experienced the additional mixing at/after the RGB bump, and establish their precise 
evolutionary stage. Being the spectra obtained with different instruments and configurations, 
we need to use different indicators for the C abundance and its isotopic ratio. 
For the ELODIE spectra and for those stars observed with the UVES grating CD\#2,  
C abundance and $^{12}C$/$^{13}C$ ratio were estimated by spectral synthesis of the 
$^{2}\Delta$--$^{2}\Pi$ band of CH (the G band) at $\sim$4310 $\mathring{A}$.
For the stars for which the grating CD\#2 is not available, we used the bluest 
portion of the spectrum (obtained with the CD\#1 grism), covering the 
spectral region of 3000--3800 $\mathring{A}$. For these stars C was obtained 
from the $^{2}\Sigma$--$^{2}\Pi$ band of CH at 3143 $\mathring{A}$ and from other 
CH transitions in the range between 3088 and 3115 $\mathring{A}$. These regions have been used 
also to measure the $^{12}C$/$^{13}C$ ratio. The atomic and molecular linelists were taken from the 
Kurucz compilation \footnote{http://kurucz.harvard.edu/}. For the CH transitions, the 
log~{\sl gf} by the Kurucz database were revised downward by 0.3 dex, in order to well reproduce the 
solar-flux spectrum by \citet{neckel84} with the C abundance by \citet{caffau10}; similar corrections 
were applied also by other authors, see e.g.. \citet{bonifacio98} and \citet{lucatello03}.
As a first guess we derived C by assuming the solar $^{12}C$/$^{13}C$ ratio; then, we computed 
synthetic spectra varying the $^{12}C$/$^{13}C$ ratio, keeping C fixed. The procedure was
repeated until the convergence within a tolerance of 0.1~dex in the C abundance. 
Only upper limits can be inferred for several stars in our sample, due to the low intensity of the 
$^{13}CH$ transitions (only for the most metal-poor star in the sample, namely CD~--$30^{\circ}$0298, 
we cannot provide a reliable and strict upper limit).

\begin{table*}
\begin{minipage}{120mm}
\caption{Identification numbers, adopted instrument, temperatures, gravities, microturbulent velocities,
[Fe/H] and A(Li) abundances,  $^{12}C/^{13}C$  isotopic ratio and SNR around the Li line, for the observed stars. 
} \label{dat}
\begin{tabular}{lccccccccc}
\hline
ID &  Instrument&   $T_{eff}$ & log~g & $v_t$ & [Fe/H]   & A(Li)  & $^{12}C/^{13}C$ & SNR\\
\hline
   &      &   (K)  &   &  (km/s)   &  (dex)  &  (dex)  &   &  \\
\hline   
HD~2665         	& ELODIE  &   4950     &   2.20   &  1.80	&  --2.18  &	0.95	 &    40     &   250  \\
HD~6755         	& ELODIE  &   5100     &   2.30   &  1.50	&  --1.57  &	1.01	 & $>$20     &   110  \\
HD~26169         	& UVES    &   5050     &   1.90   &  1.50	&  --2.43  &	0.99	 & $>$30     &   320  \\
HD~27928        	& UVES    &   5000     &   2.00   &  1.50	&  --2.33  &	1.04	 & $>$30     &   370 \\
HD~45282        	& ELODIE  &   5200     &   2.80   &  1.50	&  --1.56  &	1.01	 & $>$30     &   230  \\
HD~87140        	& ELODIE  &   5050     &   2.30   &  1.50	&  --1.83  &	0.92	 &    35     &   150  \\
HD~111721       	& UVES    &   4950     &   2.30   &  1.30	&  --1.43  &	0.98	 &    40     &   370  \\
HD~128279       	& UVES    &   5300     &   2.60   &  1.70	&  --2.21  &	1.02	 & $>$35     &   280  \\
HD~175305       	& ELODIE  &   5050     &   2.40   &  1.50	&  --1.45  &	1.07	 &    30     &   180    \\
HD~200654       	& UVES    &   5150     &   2.10   &  1.50	&  --2.96  &	0.98	 & $>$35     &   290  \\
HD~211998       	& UVES    &   5200     &   2.90   &  1.20	&  --1.59  &	0.97	 &    40     &   540  \\
HD~218857       	& UVES    &   5000     &   1.90   &  1.30	&  --2.05  &	0.92	 & $>$25     &   330   \\
HD~274939       	& UVES    &   5000     &   2.00   &  1.50	&  --1.74  &	0.96	 & $>$20     &   310  \\
BD~--$01^{\circ}$2582   & UVES    &   5000     &   1.90   &  1.10	&  --2.39  &	0.82	 &    40     &   280  \\
BD~+$23^{\circ}$3130    & ELODIE  &   5200     &   2.20   &  1.50	&  --2.59  &	0.92	 & $>$15     &   170  \\
CD~--$24^{\circ}$1782   & UVES    &   5000     &   2.10   &  0.80	&  --2.91  &	0.96	 & $>$20     &   280  \\
CD~--$30^{\circ}$0298   & UVES    &   5100     &   2.10   &  1.80	&  --3.40  &	0.99	 &   ---     &   330  \\
\hline
\end{tabular}
\end{minipage}
\end{table*}


\subsection{Lithium abundance from excitation temperatures}

We performed a fully spectroscopic analysis of the target stars, to derive their atmospherical parameters. 
They were computed  by imposing no trend between neutral iron lines 
abundances and excitation potential $\chi$ (to constrain $T_{eff}$), no trend 
between neutral iron lines abundances and the reduced equivalent width 
$\log{EW/\lambda}$ (to constrain the microturbulent velocity) and 
the same abundance (within the quoted uncertainties) from neutral and single 
ionized iron lines (to constrain the gravity).

We noted that the iron lines with $\chi<\sim$1 eV exhibit abundances systematically higher 
than the other lines for stars with metallicity lower than --2.5 dex. Even if the adopted $T_{eff}$ 
provides no dependence between Fe~I abundance and $\chi$ in the range of $\chi\sim$1--5 eV, the 
low-$\chi$ lines give abundances higher by $\sim$0.3 dex. We excluded from the analysis the 
Fe~I lines with $\chi<$1 eV, because their inclusion can force a too low $T_{eff}$ 
(with a reduction of about 200 K), in order to erase a spurious slope. For more metal-rich stars, 
this discrepancy is reduced or totally erased.
This effect was already noted in other studies of metal-poor stars \citep{norris01,carretta02,cayrel04}
and probably ascribable to some inadequacies of the model atmospheres 
based on the assumptions of 1-dimensional geometry and/or LTE:
in fact, the low-$\chi$ lines are typically formed in the outermost layers of the photosphere, 
that are exposed to the UV radiation coming from the deep layers (thus, the non-LTE effects are more pronounced in 
metal-poor stars because of the low opacity and high transparency of the photosphere).

The derived iron abundances range from [Fe/H]=~--3.40 dex to [Fe/H]=~--1.43 dex.
The average Li abundance inferred from these 17 giants turns out to be A(Li)=~0.97 dex ($\sigma$=~0.06 dex).

\begin{figure}
\includegraphics[width=84mm]{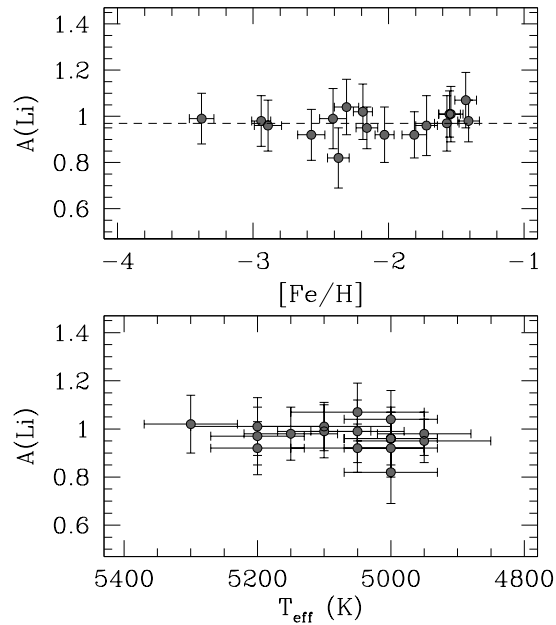}
\caption{Behaviour of the Li abundance as a function of the iron abundance 
(upper panel) and of the temperature (lower panel). The dashed line shown in the upper 
panel represents the average A(Li) of the field stars of the sample. }
\label{res}
\end{figure}

Figure~\ref{res} shows the behaviour of the sample stars as a function of [Fe/H] and 
$T_{eff}$ (upper and lower panel, respectively). We checked for the occurrence of 
slopes in these two planes, performing least-square fits by considering 
the errors in both variables 
\citep[following the approach by][]{press92}. 
In the A(Li)--[Fe/H] plane we derived a slope of 0.018 dex/dex (with an error of 0.044 dex/dex), while 
in the A(Li)--$T_{eff}$ we find a slope of 0.013 dex/100 K (with an error of 0.028 dex/100 K). 

Eight stars in our sample are in common with \citet{gratton00}. 
We find a difference in the adopted temperatures of $T_{eff}$--$T_{eff}^{Gratton00}$=~--107 K 
($\sigma$=~41 K), probably due to their $T_{eff}$ scale,  
based on the grids of synthetic colors computed by R. L. Kurucz by means of the ATLAS9 model 
atmospheres with the inclusion of the approximate overshooting and adopting the old 
set of Opacity Distribution Function \citep[we refer the reader to]
[for details about the differences between the two set of ATLAS9 models]{castelli04}.
The mean A(Li) difference for the stars in common is  
A(Li)--${\rm A(Li)}^{Gratton00}$=~--0.19 dex ($\sigma$=~0.07 dex), partially due to the 
different $T_{eff}$ scales. The residual discrepancy most likely stems from the 
adoption of overshooting model atmospheres, that provide 
higher Li abundances, as pointed out by \citet{molaro95}.

Ten stars are in common with \citet{gp06}; we find average differences 
$T_{eff}$--$T_{eff}^{Garcia-Perez06}$=~--80 K ($\sigma$=~84 K) and 
A(Li)--A(Li)$^{Garcia-Perez06}$=~--0.11 dex ($\sigma$=~0.11 dex).

\begin{figure}
\includegraphics[width=84mm]{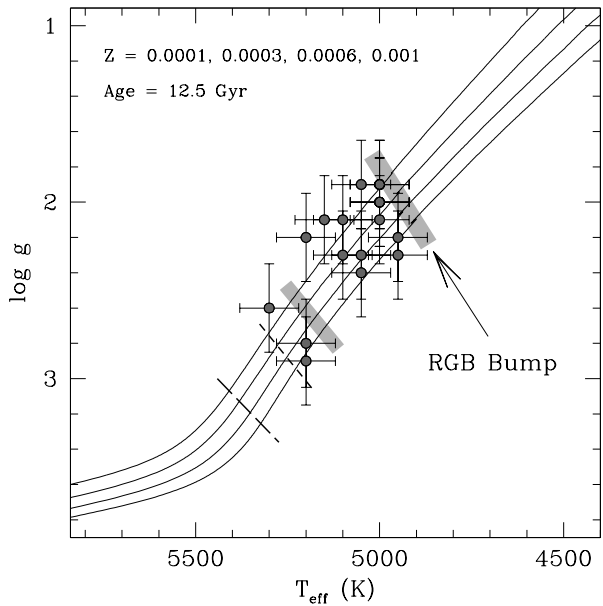}
\caption{Position of the target stars (grey points) in the $T_{eff}$--log~g diagram, 
compared with the $\alpha$-enhanced theoretical isochrones from the BaSTI database, 
for a reference age of 12.5 Gyr and different metallicities (namely Z=0.0001, 0.0003, 0.0006 and 
0.001). The isochrones are plotted only up to the tip of the RGB, for sake of clarity. 
The location of the RGB bump is marked as a grey shaded region. The fainter 
shaded region marks the boundary beyond which A(Li) stops decreasing, i.e. attains its 
final value along the RGB. 
We also display two lines corresponding to A(Li) increased by 
0.01~dex and 0.10~dex, (short-dashed and long-dashed, respectively).  
}
\label{iso}
\end{figure}

Figure~\ref{iso} shows the position of the target stars in the $T_{eff}$--log~g plane, 
together with theoretical isochrones by \citet{pietrinferni06}, computed with an age of 
12.5 Gyr and metallicities Z=0.0001, 0.0003, 0.0006 and 0.001. These isochrones were computed 
with the same input physics, code, metal distribution and $\Delta$Y/$\Delta$Z of our calculations.
The position of the RGB bump is marked for reference as a grey shaded region. 
All the stars are located (within the 
uncertainties in the atmospherical parameters) below the RGB bump level, as confirmed also by the 
$^{12}C/^{13}C$ ratio higher than $\sim$15--20. In fact, the occurrence of the extra-mixing episode 
after the RGB bump decreases dramatically the $^{12}C/^{13}C$ ratio, reaching  values lower than 10 
\citep[see for instance][]{gratton00,spite06}. This confirms, together with the homogeneous 
distribution of A(Li), that all these stars have not yet experienced the extra-mixing episode.
The figure also shows that essentially 
all stars in our sample have attained the final surface A(Li) values, before the onset of the 
post RGB bump extra mixing.  

\subsection{Lithium abundance from Infrared Flux Method temperatures}

As a sanity check to assess the robustness of the average A(Li) value we 
repeated the analysis of the target stars with other $T_{eff}$ scales,  
the temperature being the most crucial parameter in the derivation of A(Li).
We infer $T_{eff}$ in our targets by means of suitable transformations between 
dereddened broad-band colors and effective temperatures 
obtained through the classical Infrared Flux Method \citep[IRFM;][]{blackwell}.
Several $T_{eff}$ scales based on this technique are available in literature 
\citep[e.g.][]{monte98,alonso99,ramirez05,ghb09,casa10}, that make use 
of different model 
atmospheres, photometric systems and recipes for the absolute calibrations and zero-points.
The comparison between the various IRFM scales is beyond the purpose of this work 
and we focus our attention only on the widely used in literature calibration by 
\citet[][hereafter A99]{alonso99} 
and the recent one by \citet[][hereafter G09]{ghb09}, 
based on the Two Micron All Sky Survey 2MASS photometric data to compute the 
infrared monochromatic flux of the targets.

In order to have a homogeneous set of magnitudes, we adopted for all the field stars 
the J and $K_s$ magnitudes available in the final release of the 2MASS catalogue \citep{skrut}. 
Note that the 2MASS magnitudes of the targets were transformed to the photometric 
system of the Telescope Carlos Sanchez (where the calibration by A99
is defined) by means of the relations by \citet{carpenter} and \citet{alonso98}.

The $(J-K)_0$ colors were obtained by adopting the extinction coefficients by \citet{mccall04} 
and the color excess E(B-V) from the infrared dust maps by \citet{schlegel},  
corrected following the prescriptions by \citet{boni_monai}.

For some stars, E(B-V) is larger than 0.4 mag and 
the derived $T_{eff}$ is incompatible with the evolutionary stage of the targets 
and with the spectroscopic $T_{eff}$. 
For these stars we computed the color excess E(B-V) from the EW of 
the interstellar doublet Na~I lines (at 5890 and 5896 $\mathring{A}$),
through the calibration by \citet{munari97}.

Gravities and microturbulent velocities have been derived spectroscopically, as described 
in the previous Section.

The differences between the photometric and spectroscopic scales are equal to 
$T_{eff}^{GB}$-$T_{eff}^{spec}$=~+102 K ($\sigma$=~55 K) and 
$T_{eff}^{A99}$-$T_{eff}^{spec}$=~+4 K ($\sigma$=~57 K). This difference 
is fully consistent with the intrinsic difference between the two scales
(see the discussion in G09). 
Consequently, we derived an average A(Li) of 0.97 dex ($\sigma$=~0.07 dex) and 
1.07 dex ($\sigma$=~0.07 dex) when $T_{eff}$ by A99 and G09
are adopted, respectively.

Thus, the spectroscopic $T_{eff}$ and those obtained with the 
A99 calibration can be considered on the same scale, while the scale 
by G09 is slightly hotter, providing Li abundances $\sim$0.1 dex higher.

Finally, we note that 
the general agreement between spectroscopic and photometric $T_{eff}$ 
(apart from the different zero-points) seems to indicate that no relevant 
departures from the LTE condition occurs, at least when the low-$\chi$ lines 
are excluded for the most metal-poor stars.

To date, there is no general consensus about the magnitude of non-LTE corrections  
for iron, due to the incompleteness of the Fe model atom and the 
uncertainty about the rate of collision with the hydrogen atoms. Adopting several recipes 
for the calibration of the $S_H$ parameter, the scaling-factor 
to correct the 
H I collision rate provided by the Steeenbock \& Holweger 
generalisation \citep{SH84} of the Drawin formula \citep{D68,D69}, 
different authors provide different 
non LTE corrections for iron 
\citep[see for instance][providing very diferent results]{gratton99,gehren04}.
 
\citet{mashonkina} analyzed a metal-poor giant star (slightly colder than 
those discussed here) by considering the LTE case and the non LTE case with different 
efficiencies for the collision rate, and found that in case of $S_H$=0.1 (corresponding 
to a low efficiency  of collisions with H I) a satisfactory excitation equilibrium is reached only by 
decreasing the temperature of $\sim$80~K.

Recently \citet{barklem} have convincingly shown that 
the Drawin formula lacks the necessary physical  ingredients
to properly capture the quantum mechanical processes involved in
excitation and ionisation, through collisions with hydrogen atoms.
In fact in the cases for which such quantum mechanical computations
have been performed the results differed by several orders 
of magnitude with respect to the predictions based on the 
Drawin formula.
This casts some serious doubts on the results of any NLTE computation
which depends sensitively on the use of the Drawin formula
and a suitable scaling factor.

In our case the good agreement between our photometric and spectroscopic $T_{eff}$, 
regardless of the metallicity, suggests that no significant non-LTE effects on iron
are at work, at least for the lines of high excitation.

\section{Theoretical analysis}

Our theoretical analysis is based on a reference grid of 
stellar evolution models calculated with and without including the effect of atomic diffusion.  
More details about the code and input physics are in \citet{pietrinferni06} and \citet{muc} . 
We have considered the $\alpha$-enhanced ([$\alpha$/Fe]$\sim$0.4) 
metal mixture of \citet{pietrinferni06} and pairs of He and metal mass fractions 
(Y,Z) equal to (0.245,0.00001),(0.245,0.0001),(0.245,0.0003), \- (0.248,0.002),(0.251,0.004)   
that correspond to [Fe/H]=$-3.62, -2.62, -2.14, -1.31$, and $-1.01$,  
respectively. The adopted $\Delta$Y/$\Delta$Z=1.4 is the same as in \citep{pietrinferni06}.  
For each metallicity we calculated the evolution of one stellar mass  
whose age along the RGB is equal to $\sim$12.5~Gyr, and followed its evolution from the pre-MS to luminosities beyond the RGB bump.  
The range of mass values is comprised between $\sim$0.80 and $\sim$0.86$M_{\odot}$, increasing with increasing Z. 
The initial A(Li) for all metallicities was fixed at the BBN predicted value of 2.72~dex. We have also calculated 
models with an initial Li abundance increased or decreased 
by one order of magnitude and found -- as expected -- that the amount of depletion after the first dredge-up ($\Delta$(Li)) 
is insensitive to the exact value of the initial A(Li).

\begin{figure}
\includegraphics[width=84mm]{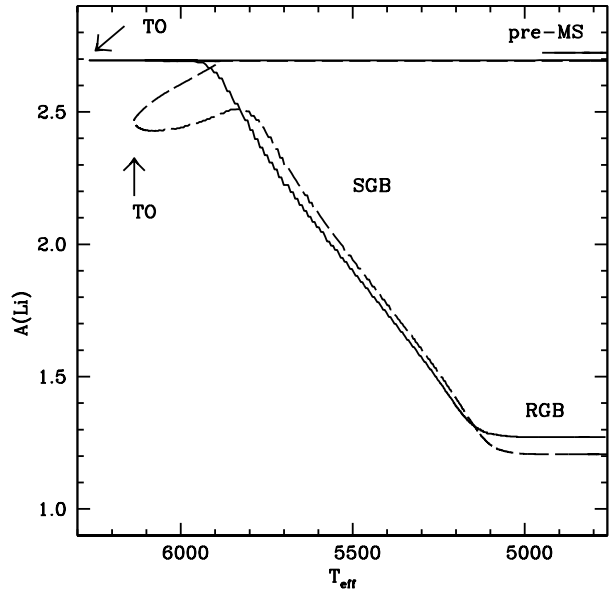}
\caption{Evolution of the surface A(Li) as a function of $T_{eff}$ for 0.855$M_{\odot}$, $\alpha$-enhanced, [Fe/H]=$-$1.01 
stellar models without (solid line) and with (dashed line) the inclusion of atomic diffusion. The pre-MS, turn off (TO), 
subgiant branch (SGB) and RGB regions are marked in the figure.}
\label{Lievo}
\end{figure}

Figure~\ref{Lievo} displays the evolution of the surface Li abundance of models   
with initial [Fe/H]=$-$1.01. 
The evolution without diffusion displays a 0.03~dex pre-MS depletion, followed by the MS evolution 
with unchanged surface A(Li). Dilution due to the deepening of the convective envelope at the end of the MS 
starts at $T_{eff}\sim$5900~K, and continues until the first dredge-up is completed, 
at $T_{eff}\sim$5100~K, when the model is already evolving along the RGB. 
As for the evolution with fully efficient atomic diffusion, the surface Li decreases during the MS and 
reaches a minimum around the turn off. After this point A(Li) at first increases, due to the deepening convection  
that returns to the surface part of the Li diffused outside the envelope along the MS. After a local maximum at 
$T_{eff}\sim$5850~K, A(Li) starts to drop, for the convective envelope reaches regions where Li has been burned 
\citep[see, e.g.][]{del90,sw01}. The surface Li abundance at the end of the dredge-up is only  0.07~dex lower 
when diffusion is fully efficient. 

Table~\ref{depl} displays $\Delta$(Li) (in dex) for the metallicity range 
spanned by our calculations, and an age of 12.5~Gyr. An important point to notice is that the [Fe/H] values 
reported in the first column are the initial values. In models without diffusion [Fe/H] at the surface 
stays constant all along the 
MS and is negligibly affected by the increase of surface He (and associate decrease of H) after 
the first dredge-up. 
However, when diffusion is fully efficient, it decreases during the MS, because of the sinking of Fe 
and increase of H in the envelope. It is only after the first dredge-up that [Fe/H] is restored to within 
$\sim$0.02-0.03~dex of the initial value. Working with lower RGB stars removes also the ambiguity 
between initial and actual [Fe/H] when diffusion is efficient. 

Overall, the total Li depletion from the pre-MS to the RGB shows a mild dependence on the initial 
metallicity. Metal rich models 
display a more efficient depletion, due essentially to deeper convective envelopes along the RGB. This is in agreement 
with previous results by \citet{del90} and \citet{sb99}. The 
values of $\Delta$(Li) cover a range of $\sim$0.15~dex over the [Fe/H] interval spanned by our calculations. The 
variation of  $\Delta$(Li) is non-linear, with an increase of $\Delta$(Li)/$\Delta$[Fe/H] with 
increasing [Fe/H].

\begin{table}
\caption{Li abundance depletion ($\Delta$(Li), in dex) along the RGB for the labelled 
[Fe/H] values. We report the results for models without diffusion, with the inclusion of diffusion, 
and without diffusion but accounting for overshooting below the envelope convection, respectively  
(see text for details).} \label{depl}
\begin{tabular}{lccc}
\hline
 [Fe/H]  & $\Delta$(Li) (no diff) &  $\Delta$(Li) (diff)  & $\Delta$(Li) (oversh)\\
\hline   
$-$3.62 &  1.28 &  1.35 &  1.29\\ 
$-$2.62 &  1.30 &  1.37 &  1.31\\
$-$2.14 &  1.33 &  1.40 &  1.34\\
$-$1.31 &  1.40 &  1.46 &  1.41\\
$-$1.01 &  1.44 &  1.51 &  1.50\\
\hline
\end{tabular}
\end{table}

We have also calculated models without diffusion but including -- starting after the end of the MS -- 
overshooting below the convective envelope. The extension of the overshooting region has been parametrized 
in terms of the pressure scale height ($H_p$) at the Schwarzschild boundary. We assume instantaneous mixing and 
radiative temperature gradient in the overshooting region \citep[see also][for similar calculations but 
a much larger overshooting region]{del90}.
The need for these additional calculations stems from 
the results of analyses of the RGB bump luminosity in Galactic globular clusters. The predicted bump luminosities from 
models calculated with the same code and input physics adopted here, 
are too bright compared to observations, by $\sim$0.2~mag on average
\citep[see][and references therein]{cas11} although there is a hint 
of an increase of the discrepancy towards the lowest metallicities. In these additional calculations 
we have considered an overshooting distance equal to 0.35~$H_p$ for the metal poor models up to [Fe/H]=$-$2.14. This 
overshooting length lowers the RGB bump magnitude by $\sim$0.40~mag.
For models with  [Fe/H]=$-$1.31 and $-$1.01 we have considered an overshooting distance equal to 0.10~$H_p$, that lowers 
the bump level by $\sim$0.2~mag.
The corresponding values of $\Delta$(Li) along the lower RGB are reported in Table~\ref{depl}. 
The effect of overshooting from the convective envelope along the RGB -- with our calibration -- is negligible 
($\Delta$(Li) increases by only 0.01~dex) with the exclusion of the model with [Fe/H]=~-1.01 dex for which 
the difference is of 0.06 dex. 
We have verified -- with calculations for [Fe/H]=$-$3.32, $-$1.31 and $-$1.01 -- 
that the effects of diffusion and overshooting (at least for 
our choice of the overshooting extension) on  $\Delta$(Li) are additive, when we compute models including both effects.

In addition, we have explored the effect of varying age, initial He content and mixing length, by calculating selected 
models at [Fe/H]=$-$2.14.  
We found that increasing the age by 2 Gyr above our reference value of 12.5~Gyr decreases $\Delta$(Li) 
by 0.02~dex. A decrease  
of the mixing length by 0.2$H_p$ below our solar calibrated value (2.01$H_p$) leaves $\Delta$(Li) unchanged, 
while an increase of the initial He mass fraction Y by 0.01 increases the depletion by 0.01~dex only.

\subsection{Estimate of the primordial Li abundance in Population II stars}

As we have discussed in the section on the observational data, the measured values of the$^{12}C/^{13}C$ ratio and 
the position in the $T_{eff}$--log~g diagram place our star sample in the lower RGB phase, after the completion 
of the first dredge-up and before the onset of the mixing episode at the RGB bump luminosity. 
We have determined their primordial A(Li) -- that we denote as ${\rm A(Li)}_0$, to compare with 
the ${\rm A(Li)}_0$ predicted by the BBN nucleosynthesis  -- for each of the three adopted 
$T_{eff}$ scales, making use of a Monte Carlo technique, as detailed below. 

For each temperature scale 
we have first fixed the set of theoretical models. From the chosen set, 
we have produced a large (500 objects) sample of lower RGB synthetic A(Li) abundances, 
by drawing random values (flat distribution, consistent with the distribution in the observed sample) of [Fe/H] 
within the range spanned by our models, and determining the corresponding depletion $\Delta$(Li)  
by linear interpolation among the values listed in Table~\ref{depl}. 

\begin{table}
\caption{Estimates of the cosmological Li abundance (${\rm A(Li)}_0$) 
in our sample of lower RGB Halo stars. We have considered three sets of stellar 
evolution models with different assumptions about the element transport mechanisms, 
and three $T_{eff}$ scales for the A(Li) determinations (see text for details).} \label{initial}
\begin{tabular}{lccc}
\hline
 Models  & ${\rm A(Li)}_0$ (A99) &${\rm A(Li)}_0$ (G09) & ${\rm A(Li)}_0$ (spect)\\
\hline   
Standard     &  2.28 &  2.39 &  2.30\\ 
Diffusion    &  2.35 &  2.46 &  2.37\\
Overshooting &  2.29 &  2.40 &  2.31\\
\hline
\end{tabular}
\end{table}

\begin{figure}
\includegraphics[width=84mm]{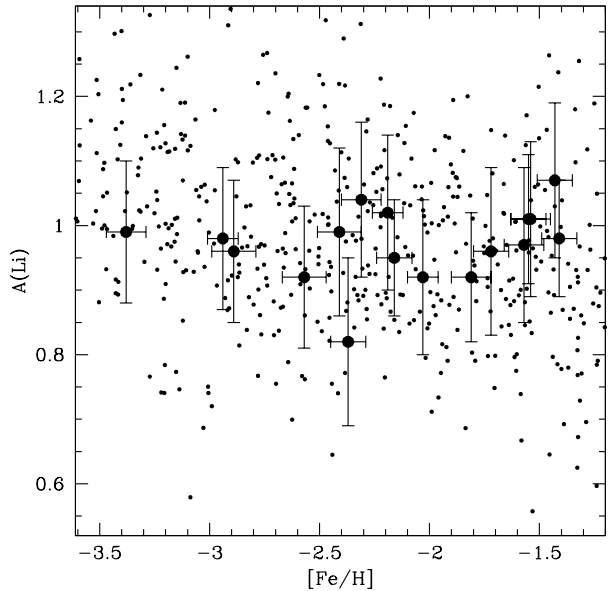}
\caption{Comparison between a synthetic sample of A(Li) values from standard models and  
${\rm A(Li)}_0$=2.30, and the observed A(Li) obtained employing the  
spectroscopic $T_{eff}$ scale (see text for details).}
\label{synthe}
\end{figure}

The value of ${\rm A(Li)}_0$ has then been determined as follows. To each synthetic $\Delta$(Li) we 
have added a constant that corresponds to a trial value for the initial Li abundance of the sample. 
Each element of the resulting ([Fe/H], A(Li)) synthetic sample has been then perturbed by a 
random Gaussian error. These error values have been calculated taking the mean 
$<\sigma_{obs}>$ spectroscopic error from the observations, perturbed by  
a Gaussian random distribution with standard deviation equal to the 
1$\sigma$ dispersion of the spectroscopic errors around the value $<\sigma_{obs}>$.
The initial Li abundance of the synthetic sample is then varied until the mean A(Li) of the synthetic sample 
matches the observed value. The corresponding 
initial Li-abundance is considered to be the best estimate of ${\rm A(Li)}_0$.
Finally, we have performed a Kolmogorov-Smirnov test to compare the synthetic distribution 
calculated with the best estimate of ${\rm A(Li)}_0$ with the observed one, and 
verified that, for all our choices of models and $T_{eff}$ scale, the probability $P$  
that the two samples are not drawn from the same distribution is well below the 95\% threshold.
A comparison between synthetic and observed samples for a particular choice of the 
$T_{eff}$ scale and stellar models is displayed in Fig.~\ref{synthe}. 

Our final estimates of ${\rm A(Li)}_0$ are reported in Table~\ref{initial}.
Models that include atomic diffusion without overshooting predict ${\rm A(Li)}_0$
larger by 0.07 dex with respect to the standard models, while models including 
only overshooting lead to Li abundance larger by only 0.01 dex when compared with the standard 
case  (at least for the metallicity range covered by our targets). Inclusion of both diffusion and overshooting would increase
${\rm A(Li)}_0$ of 0.08 dex when compared with the standard models.

\section{Li abundances in lower RGB globular cluster stars}

It is interesting to investigate also the surface A(Li) in lower RGB stars belonging to Galactic 
globular clusters, in comparison with the values obtained for field stars. 
A note of caution is however warranted.  
The estimate and interpretation of the initial Li abundance of Galactic globular clusters in terms of Halo 
primordial Li should be considered with caution, 
because of the occurrence of self-enrichment processes during the 
formation/evolution of the clusters, that are potentially able to affect the measured A(Li).  
A point to recall is that, in light of the currently accepted 
self-enrichment scenario for the evolution of the 
globular clusters, these systems lose a relevant fraction of their first generation stars -- whose chemical 
abundance pattern is expected to be consistent with the field Halo population -- during their 
early evolution. Spectroscopic studies \citep[see e.g.][]{carretta10,martell10} show that the large majority 
of field Halo stars have in fact an abundance pattern consistent with those of first generation stars in globulars, 
hence both kind of objects are compatible in terms of 
A(Li). Only a very marginal fraction (less than 2\%) of field halo stars is compatible with 
second generation cluster stars -- that show the well known CN and ONa anticorrelation and have 
Li abundances potentially modified by nuclear processing.

Lithium abundances are available (from the analysis of dwarfs and/or 
giants) for a few of the closest clusters. 
To date, NGC~6752 \citep{shen} and 47 Tuc \citep{dorazi10} exhibit clear signatures 
of intrinsic star-to-star dispersion in A(Li), whereas M4 show a very high degree of homogeneity 
in A(Li) at a given evolutionary phase \citep{muc}. NGC~6397 shows also similar homogeneous A(Li) abundances, 
with the detection of only 3 (out of 100) dwarf stars with A(Li)$<$2 \citep{lind}.

The interpretation of these results is beyond the purpose of this work, and we
discuss here only the Li abundance in giants stars of these clusters (namely NGC~6397, NGC~6752 and M4) 
for which high-resolution spectra of giant stars are available, to provide a comparison 
with the Li abundances inferred in field RGB stars.

\subsection{NGC~6397}

We have analyzed 45 RGB stars (below the RGB bump level) in NGC~6397, retrieved from 
the dataset of GIRAFFE/FLAMES spectra discussed by \citet{lind}. All the observations 
are performed with the HR13 and HR15 gratings. 
Due to the limited number of available iron lines, a full spectroscopic analysis cannot be 
performed and we discuss only the Li abundances inferred by employing the photometric 
$T_{eff}$ scales by A99 and G09. We employed the $(J-K)_0$ color, 
obtained from 2MASS photometry, corrected for reddening using the value of E(B-V) 
provided by \citet{f99}. 
To reduce the scatter in the derived A(Li) distribution due to the photometric errors, we 
projected each star along the fiducial line of the RGB sequence in the K--(J-K) plane 
(note that this technique, obviously applicable to stars members of a stellar cluster and not 
to field stars, is correct under the assumption that 
the broadening of the RGB is due to photometric errors only).

In their original work \citet{lind} derive $T_{eff}$ values by using the synthetic Str\"omgren {\sl (b-y)} 
colors computed by \citet{onehag09} employing MARCS model atmospheres, and find an average 
A(Li)=~1.13 dex ($\sigma$=~0.09 dex) for the lower RGB sample. 
We find that the G09 and \citet{lind} $T_{eff}$ scales well match, 
with an average difference $T_{eff}^{GB}-T_{eff}^{Lind09}$=~--13 K ($\sigma$=~15 K), 
while the difference with the temperatures by A99 is 
$T_{eff}^{Alo}-T_{eff}^{Lind09}$=~--117 K ($\sigma$=~14 K).
The iron content is [Fe/H]=~-2.08 dex ($\sigma$=0.09 dex) with the G09 scale and 
-2.12 dex ($\sigma$=0.08 dex) with the A99 scale.
We find an average A(Li)=1.00~dex ($\sigma$=~0.09 dex) with the A99 scale and 
of A(Li)=1.09 dex ($\sigma$=~0.10 dex) with the G09 scale. 

By interpolating in [Fe/H] among the values of Tab.~2, 
we derive an estimate for ${\rm A(Li)}_0$ of 2.33--2.42~dex with the model without diffusion 
(A99 and GH09 scales respectively) and 2.40--2.49~dex with the model with 
diffusion.

\subsection{NGC~6752}

Twenty-one RGB stars were observed with UVES within the ESO Large Program 65.L-0165 (PI: Grundahl). 
Concerning the Li abundance, only the qualitative behaviour of the observed EWs as a function 
of V magnitude was discussed  \citep{grund}, without the explicit determination of the 
Li abundance. 
Here we consider the 12 stars fainter than the RGB bump, for which the Li line 
is clearly detectable. We analyzed the spectra obtained with the CD\#3 cross-disperser, 
deriving $T_{eff}$ both spectroscopically and photometrically, adopting the 
A99 and G09 calibrations for the $(J-K)_0$ colour. 
The J and $K_s$ magnitudes are from the 2MASS database, corrected for reddening 
using the E(B-V) value by \citet{f99}. Also for this cluster, we minimized the scatter in the final 
A(Li) values by employing $(J-K)_0$ 
colours obtained projecting the position of each star along the RGB fiducial line.

\citet{grund} derived the $T_{eff}$ for their targets by means of the 
A99 calibration for the Str\"omgren {\sl (b-y)} index. 
We find a reasonable consistency with our spectroscopic and photometric 
$T_{eff}$ on the A99 scale (as for the case of field stars)  
with average differences $T_{eff}^{Spec}-T_{eff}^{Grundahl02}$=~--35 K 
($\sigma$=~45 K) and $T_{eff}^{Alo}-T_{eff}^{Grundahl02}$=~--43 K ($\sigma$=~25 K), 
respectively. The offset with the $T_{eff}$ by G09 is 
of $T_{eff}^{GB}-T_{eff}^{Grundahl02}$=~+65 K ($\sigma$=~26 K).

The average iron content is [Fe/H]=~--1.68 dex ($\sigma$=0.07 dex), when the 
spectroscopic temperatures are used and different by only a few hundredths dex 
when the photometric $T_{eff}$ are adopted.
The average A(Li) is 0.83 dex ($\sigma$=~0.15 dex) with the 
spectroscopic temperatures (almost the same abundance is obtained with the A99 scale, 
while the G09 scale provides A(Li)=~0.93 dex, $\sigma$=~0.15 dex).
Three stars display A(Li)$\sim$0.5-0.6 dex, lower by $\sim$0.3-0.4 dex compared to 
the other stars in the sample. The remaining 9 stars provide an average Li abundance 
A(Li)=~0.91 dex ($\sigma$=~0.07 dex). The presence of these three Li-poor stars 
\citep[already identified by][]{grund} is probably ascribable 
to the lithium variation in the cluster, as testified by the Li-Na anticorrelation \citep{pasquini05} 
and Li-O correlation \citep{shen} . These stars could belong to the second generation.

By employing the $\Delta$(Li) values listed in Tab.~2, the models without and with the inclusion 
of the diffusion provide ${\rm A(Li)}_0$ estimates of 2.29--2.35, 
2.18--2.24 and 2.19--2.25 dex, by adopting, respectively, the $T_{eff}$ scale 
from GH09, A99 and from the excitation equilibrium. 
When we exclude the 3 Li-poor stars, the derived ${\rm A(Li)}_0$ increase by 0.08 dex.

\subsection{M4} 

We summarized briefly the results discussed in \citet{muc} about the metal-rich 
cluster M4, from the analysis of a sample of GIRAFFE spectra. 
The differential reddening that affects the field of view of M4 makes 
uncertain the abundances derived directly by adopting the photometric $T_{eff}$, 
due to the residual of the  differential reddening correction. 
The atmospheric parameters for the RGB stars in the sample were therefore  
derived by projecting the position of each star along the stellar isochrone the best fit the 
observed colour-magnitude-diagram. 
This $T_{eff}$ scale is in nice agreement  with the 
spectroscopic $T_{eff}$ values inferred by \citet{marino} for the stars in common.

The derived Li abundance in lower RGB stars of M4 is A(Li)=0.92~dex, that leads to   
${\rm A(Li)_0}$=2.35~dex (models without diffusion) and 2.40~dex 
(models with diffusion).

\section{Summary and conclusions} 

We have discussed the use of Li abundances measured in Population~II 
lower RGB stars as an independent, reliable and robust diagnostic of the initial Li abundance in the Galactic Halo. 
Surface abundances in giant stars fainter than the RGB bump are sensitive to the total Li content left 
at the end of the MS phase, and are very weakly affected by atomic diffusion during MS. 
Also, the predicted A(Li) in these objects is basically insensitive to 
the mixing length calibration, the precise stellar ages, initial He abundances, 
and realistic estimates of the overshooting extension 
below the Schwarzschild boundary of the convective envelope. 
Chemical abundance measurements in giant stars also suggest 
that any additional element transport along the RGB is very likely inefficient in this phase.
Overall, our analysis reveals that the predicted Li 
depletion $\Delta$(Li) along the lower RGB is robust in terms of theoretical interpretation.

The values of ${\rm A(Li)_0}$ inferred from our sample of lower RGB Halo field stars range 
from 2.28 (obtained with the A99 scale and without the inclusion of atomic diffusion) 
to 2.46 (when the G09 $T_{eff}$ scale is used, together with models including atomic diffusion).
Inclusion of overshooting from the RGB convective boundary would increase both limits by only 0.01~dex, 
while variations of age, initial He abundance and mixing length parameter provide changes of a few 
hundredths of dex or less in terms of $\Delta$(Li).
When a $T_{eff}$ scale is adopted, the effect of 
fully efficient atomic diffusion on the ${\rm A(Li)_0}$ estimate is by at most 
0.07~dex. 

The discrepancy with ${\rm A(Li)_0}$ predicted by BBN calculations thus remains, reconfirmed by the 
robustness of our estimate.
The analysis performed on the lower RGB stars in three Galactic globular clusters 
confirms similar values for ${\rm A(Li)_0}$, although in general the possible occurrence of self-enrichment processes 
in these objects has to be considered. 

There are several discussions in the literature about how to solve this discrepancy with BBN results. 
Ideas involve a first generation of stars that 
has processed and efficiently depleted lithium in a substantial fraction of the early Halo 
baryonic matter \citep{piau06}, modifications to the BBN considering the decay 
of unstable particles \citep[e.g.][]{cyburt10}, modifications to the reaction cross sections 
for the Li production during BBN \citep[e.g.][]{chak11}, or 'deep' 
turbulent mixing during the MS that connects the convective envelope to inner, Li-depleted regions 
\citep[e.g.][]{richard05, muc}. Our result for RGB stars provides an additional 
robust constraint on the efficiency of this turbulent mixing. 
Independently of its parametrization, during the MS any additional element transport 
needs to bring into the Li burning region an amount of initial 
lithium $\Delta_{burn}$(Li)=0.3-0.4~dex, in order to eliminate the discrepancy with BBN calculations 
(the turbulence model denoted as T6.25 or T6.28 by \citet{richard05} appear to burn approximately the right amount 
of Li in models with [Fe/H]=~-2.31).
The value of $\Delta_{burn}$(Li) appears to be 
roughly the same for both Halo field and globular cluster stars.

The use of lower RGB stars will allow estimates 
of the Li content in stellar populations more distant that those usually observed 
to investigate the {\sl Spite Plateau}. The obvious benefits are: (i)~ an enlarged sample 
of field stars and clusters to study the primordial Li abundance within the Galaxy, and (ii)~ 
a better chance to assess whether a ``cosmological lithium problem'' exists also in extragalactic systems 
with a different origin and a different star formation history 
\citep[see][for a first study of the initial Li in $\omega$ Cen, widely considered as the remnant of 
a dwarf galaxy accreted by the Milky Way ]{mona}. 
For instance, the current generation of high resolution spectrographs mounted on 8 metre-class telescopes  
allows to reach down to one magnitude fainter than the RGB bump level for stars in M~54 and the old stellar 
population of Sagittarius dwarf galaxy. In principle, also stars at the 
RGB bump level in old clusters belonging to the 
Large Magellanic Cloud can be reached, by using, 
for example, the VLT spectrograph X-SHOOTER; this possibility (even if expensive 
in terms of observing time due to the faintness of the targets) 
will open a new perspective in the investigation of the Li problem.
Finally, with the advent of 30 metre-class telescopes  
(as the E-ELT) spectroscopy of lower RGB stars will provide new constraints to 
the primordial Li abundance in even more distant systems. \\

The authors warmly thank the anonymous referee for his/her comments and suggestions.
PB acknowledges support from the Programme Nationale de Physique Stellaire (PNPS) and the Programme Nationale 
de Cosmologie et Galaxies (PNCG) of the Institut Nationale de Sciences de l'Universe of CNRS.

\end{document}